\documentclass[onecolumn,showpacs,amsmath,amssymb,pre]{revtex4}

\usepackage{graphicx}   % Include figure files
\usepackage{dcolumn}    % Align table columns on decimal point
\usepackage{bm}         % bold math
\begin{document}

\title[Trojan quasiparticles]
      {Trojan quasiparticles}

\author{Bettina Gertjerenken} \email{b.gertjerenken@uni-oldenburg.de}
\author{Martin Holthaus}

\affiliation{Institut f\"ur Physik, Carl von Ossietzky Universit\"at,
	D-26111 Oldenburg, Germany}

\date{July 3, 2014}

\begin{abstract}
We argue that a time-periodically driven bosonic Josephson junction supports 
stable, quasi\-particle-like collective response modes which are $N$-particle 
analogs of the nonspreading Trojan wave packets known from microwave-driven 
Rydberg atoms. Similar to their single-particle counterparts, these 
collective modes, dubbed ``flotons'', are well described by a Floquet-Mathieu 
approximation, and possess a well-defined discrete set of excitations. In 
contrast to other, ``chaotic'' modes of response, the nonheating Trojan modes 
conform to a mean-field description, and thus may be of particular interest 
for the more general question under which conditions the reduction of quantum 
$N$-particle dynamics to a strongly simplified mean-field evolution is feasible.
Our reasoning is supported by phase-space portraits which reveal the degree of 
correspondence beween the $N$-particle dynamics und the mean-field picture in 
an intuitive manner. 
\end{abstract}


% 03.75.Kk 	Dynamic properties of condensates; collective and hydrodynamic 
%		excitations, superfluid flow
% 03.75.Lm 	Tunneling, Josephson effect, Bose-Einstein condensates in 
%               periodic potentials, solitons, vortices, and topological 
%               excitations
% 03.65.Sq 	Semiclassical theories and applications
% 05.45.Mt 	Quantum chaos; semiclassical methods

\maketitle

%%%%%%%%%%%%%%%%%%%%%%%%%%%%%%%%%%%%%%%%%%%%%%%%%%%%%%%%%%%%%%%%%%%%%%%%%%%%%%%%

\section{Introduction}

The Trojan asteroids move around the sun close to the stable Lagrangian 
points $L_4$ and $L_5$ of Jupiter, sharing its orbit~\cite{RobutelSouchay10}. 
It was established only recently that also Earth has a co-orbiting Trojan 
companion, named 2010TK$_7$, whose orbit is stable over at least ten thousand 
years~\cite{ConnorsEtAl11}. In a seminal paper, Bialynicki-Birula, 
Kali\'{n}ski, and Eberly have pointed out that this classical, stable, 
periodic asteroid motion has a quantum analog in microwave-driven Rydberg atoms
which emerges when the classical Kepler frequency of the orbiting electron 
equals the microwave frequency~\cite{BialynickiBirulaEtAl94}; the nonspreading 
wave functions describing the entailing stable, though nonstationary states 
were aptly termed Trojan wave packets~\cite{KalinskiEberly96,KalinskiEberly97}. 
In a pioneering experiment, such nondispersive Trojan wave packets could be 
observed with Li Rydberg atoms for more than 
15\,000 cycles~\cite{MaedaGallagher04}; meanwhile even Trojans with principal 
quantum numbers close to $n = 600$ have been generated in a controlled 
manner~\cite{WykerEtAl12}. Theoretically, Trojan wave packets have been 
identified as Floquet states with ground state-like properties which arise 
upon quantization of a resonance zone in classical phase 
space~\cite{HenkelHolthaus92,Holthaus95,BuchleitnerEtAl02}.

Here we argue that Trojan states can also occur in quantum many-body systems, 
and then have to be regarded as stable quasiparticles which possess a 
well-defined discrete set of excitations. As a specific example, we consider a 
Bose-Einstein condensate in a double-well potential, {\em i.e.\/}, a ``bosonic 
Josephson junction''~\cite{GatiOberthaler07}, which is modulated periodically 
in time~\cite{HolthausStenholm01,JinasunderaEtAl06}. When the driving frequency
is resonant, such that it matches the slowly varying level spacing at a 
particular unperturbed state, we predict stable, nonheating, collective modes 
of response exhibiting properties typical of Trojan single-particle wave 
packets.       

We proceed as follows: In Sec.~\ref{sec:model} we introduce our model system
and its mean-field description. In Sec.~\ref{sec:resonance} we outline the 
approximate construction of a hierarchy of near-resonant $N$-particle Floquet 
states. The following phase-space analysis~\cite{MahmudEtAl05}, put forward 
in Sec.~\ref{sec:phase}, shows that the ground state of this hierarchy 
is tied to a periodic mean-field orbit in precisely the same manner as a 
Trojan single-particle wave packet is tied to a periodic solution of the 
corresponding classical equations of motion. The quasiparticle concept then 
is discussed in Sec.~\ref{sec:floton}. Although we do provide the required 
mathematical details, the main message actually is transported by phase-space 
portraits which allow one to visualize the correspondence between the 
$N$-particle- and the mean-field level. Some possible experimental
ramifications are spelled out in the final Sec.~\ref{sec:final}.

\section{Model system}
\label{sec:model}

Within the usual two-mode approximation~\cite{MilburnEtAl97,RaghavanEtAl99},
a Bose-Einstein condensate in a double well potential is described by the 
Lipkin-Meshkov-Glick Hamiltonian~\cite{LipkinEtAl65}
\begin{equation}
	H_0 = -\frac{\hbar\Omega}{2} 
	\left(a^{\phantom{\dagger}}_1 a_2^{\dagger} 
	+ a_1^{\dagger}a^{\phantom{\dagger}}_2\right) 
	+ \hbar \kappa \left(
	a_1^{\dagger} a_1^{\dagger} 
	a^{\phantom{\dagger}}_1 a^{\phantom{\dagger}}_1 
	+ a_2^{\dagger} a_2^{\dagger}
	a^{\phantom{\dagger}}_2 a^{\phantom{\dagger}}_2 \right) \; ,
\label{eq:UBJ}
\end{equation}
where $\Omega$ is the single-particle tunneling frequency and $\kappa$ 
quantifies the strength of the repulsive on-site interaction, such that 
$2\hbar\kappa$ is the repulsion energy contributed by each pair of Bose 
particles occuping the same well; the operators $a_j^{\left(\dagger\right)}$ 
(with $j=1,2)$ annihilate (create) a Boson in well~$j$. We subject this 
bosonic Josephson junction~(\ref{eq:UBJ}) to a time-periodic drive with 
amplitude $\hbar\mu_1$ and frequency $\omega$, such that the total Hamiltonian 
takes the form~\cite{HolthausStenholm01,JinasunderaEtAl06}
\begin{equation}   
	H(t) = H_0 + \hbar\mu_1 \cos(\omega t) 
	\left( a_1^{\dagger} a^{\phantom{\dagger}}_1 
	- a_2^{\dagger} a^{\phantom{\dagger}}_2 \right) \; .
\label{eq:NPH}
\end{equation}
Assuming that the double well is filled with a Bose-Einstein condensate 
consisting of $N$~particles ($N \gg 1$), the system obviously lives in an 
$(N+1)$-dimensional Hilbert space ${\mathcal H}_N$, so that its states are 
specified by vectors with $N + 1$ complex components.

Within mean-field theory, one reduces the level of complexity drastically by 
introducing an order parameter (the ``macroscopic wave function'') consisting 
of merely two complex amplitudes, whose absolute squares are supposed to 
give the fractions of particles found in the two wells~\cite{Leggett01}. 
Specifically, let $|L\rangle$ and $|R\rangle$ denote the states associated 
with the ``left'' well~$1$ and with the ``right'' well~$2$, respectively. Then 
the most general single-particle state $| \psi \rangle$ in the space spanned 
by this two-state basis can be written as 
\begin{equation}
	| \psi \rangle = \cos(\theta/2) | L \rangle
	+ \sin(\theta/2) {\rm e}^{{\rm i}\phi} | R \rangle \; ,
\label{eq:SPS}
\end{equation}	     
barring an irrelevant overall phase factor. If we assume that this 
single-particle state be ``macroscopically'' occupied by $N$~Bosons, the 
corresponding state vector in ${\mathcal H}_N$, again parametrized by the 
two angles $\theta$ and $\phi$, takes the form 
\begin{eqnarray}
	| \theta, \phi \rangle & = & \frac{1}{\sqrt{N!}}
	\left( \cos(\theta/2) a_1^{\dagger} + 
	\sin(\theta/2) {\rm e}^{{\rm i}\phi} a_2^{\dagger} \right)^N
	| {\rm vac} \rangle 
\nonumber \\ 	& = &
	\sum_{n=0}^N {\binom{N}{n}}^{1/2} \cos^n(\theta/2)
	\sin^{N-n}(\theta/2) {\rm e}^{{\rm i}(N-n)\phi} |n , N-n \rangle \; ,	
\label{eq:ACS}
\end{eqnarray}
where $| {\rm vac} \rangle$ denotes the vacuum state in ${\mathcal H}_N$,
and we have written
\begin{equation}
	| n , N - n \rangle = 
	\frac{(a_1^\dagger)^n}{\sqrt{n!}}
	\frac{(a_2^\dagger)^{N-n}}{\sqrt{(N-n)!}} | {\rm vac} \rangle 
\end{equation}
for the Fock state with $n$ particles in the left and $N-n$ particles in the 
right well. These $N$-particle states $| \theta, \phi \rangle$ are known as
``atomic coherent states''; their properties have been amply discussed in the 
literature~\cite{ArecchiEtAl72}.

Given that the mean-field order parameter is in this manner specified 
by $\theta$ and $\phi$, one needs their equations of motion. Setting 
$z = \cos\theta$, it turns out that these two equations of motion coincide with 
the Hamiltonian equations derived from the classical Hamiltonian function
\begin{equation} 
	H_{\mathrm{mf}}(z,\phi,\tau) = \frac{N\kappa}{\Omega}z^2 
	- \sqrt{1-z^2} \cos\phi 
	+ \frac{2\mu_1}{\Omega} z \cos\left(\frac{\omega}{\Omega} \tau \right) 
\label{eq:NRP}
\end{equation}
with dimensionless time $\tau = \Omega t$,
in which $z$ plays the role of a momentum and $\phi$ that of its canonically
conjugate position variable; this Hamiltonian function~(\ref{eq:NRP}) thus 
describes a nonrigid nonlinear pendulum subjected to an external time-periodic 
driving force~\cite{HolthausStenholm01,SmerziEtAl97}.

\section{Quantum resonance}
\label{sec:resonance}

Because the $N$-particle Hamiltonian~(\ref{eq:NPH}) is periodic in time,
$H(t)=H(t+T)$ with $T = 2\pi/\omega$, there exists a complete set of Floquet 
states~\cite{Shirley65}, that is, a set of solutions $|\Psi_m(t)\rangle$ 
to the time-dependent $N$-particle Schr\"odinger equation of the form 
\begin{equation}
	|\Psi_m(t)\rangle = | u_m(t) \rangle
	\exp(-\mathrm{i}\varepsilon_m t/\hbar)
	\quad ; \quad m = 0,1,\ldots,N \; ,
\label{eq:FST}
\end{equation}
with $T$-periodic Floquet functions $|u_m(t)\rangle = |u_m(t+T)\rangle$ 
which are complete in ${\mathcal H}_N$ at each instant~$t$. Such a Floquet 
state~(\ref{eq:FST}) reproduces itself after each period~$T$, up to a 
phase factor determined by its quasienergy $\varepsilon_m$, and therefore 
can be regarded as an analog of a stationary state. In this section we 
derive approximations to quite particular, ``resonant'' Floquet states 
for the driven bosonic Josephson junction~(\ref{eq:NPH}) which are 
$N$-particle counterparts of the original Trojan wave 
packets~\cite{BialynickiBirulaEtAl94,KalinskiEberly96,KalinskiEberly97}.
The analysis closely follows an early general sketch by Berman and 
Zaslavsky~\cite{BermanZaslavsky77}, and its later extension~\cite{Holthaus95}.

We denote the $N$-particle energy eigenstates of the undriven 
junction~(\ref{eq:UBJ}) as $|n\rangle$, so that $H_0|n\rangle = E_n|n\rangle$
for $n = 0,1,\ldots,N$, and assume that the eigenvalues $E_n$ are ordered
according to their magnitude. As is well known, under typical conditions
the low-energy part of the spectrum, dominated by the tunneling term, is
almost harmonic oscillator-like with a slowly decreasing level spacing,
whereas the higher-energy part, dominated by the interaction, consists of
almost degenerate doublets~\cite{GatiOberthaler07,HolthausStenholm01}. Here 
we choose the driving frequency~$\omega$ such that 
\begin{equation}
	E'_r\equiv E_{r+1} - E_r \approx \hbar\omega
\label{eq:RES}
\end{equation}
for a resonant level~$n = r$ from the former, nondegenerate part of the 
spectrum. This resonance condition~(\ref{eq:RES}) is the analog of the Trojan 
condition of equal Kepler and microwave frequency~\cite{KalinskiEberly96}.
We then make the Floquet ansatz
\begin{equation}
	| \Psi(t) \rangle = {\rm e}^{-{\rm i}\eta t/\hbar}
	\sum_n b_n | n \rangle \exp\left[
	-\frac{{\rm i}}{\hbar}\Big(E_r + (n-r)\hbar\omega\Big) t \right] \; ,
\end{equation}
assuming that the significant contributions to this superposition stem from 
states close to the resonant one. This produces the system 
\begin{equation}
	\eta b_n = \Big( E_n - E_r - (n-r)\hbar\omega \Big) b_n 
	+ 2\hbar\mu_1 \cos(\omega t) \sum_m 
	{\rm e}^{{\rm i}(n-m)\omega t} \langle n| J_z |m\rangle b_m
\end{equation}
for the coefficients $b_n$, where we have introduced the operator 
\begin{equation}
	J_z = \left( a_1^{\dagger} a^{\phantom{\dagger}}_1 
	- a_2^{\dagger} a^{\phantom{\dagger}}_2 \right)/2
\label{eq:DJZ}
\end{equation}
for the population imbalance between both wells~\cite{MilburnEtAl97,
RaghavanEtAl99}. Expanding the energy eigenvalues $E_n$  quadratically around 
$n = r$, keeping only the secular terms $m = n\pm 1$, and replacing all matrix 
elements $\langle n| J_z |n \pm 1\rangle$, somewhat arbitrarily, by the 
constant $\langle r| J_z | r-1 \rangle$, we arrive at the system
\begin{equation}
	\eta b_n = \frac{1}{2}(n-r)^2 E''_r b_n
	+ \hbar\mu_1 \langle r | J_z | r - 1 \rangle
	\Big( b_{n+1} + b_{n-1} \Big) 
\label{eq:NNS}
\end{equation}	
which couples $b_n$ to its nearest neighbors $b_{n\pm 1}$ only. When 
representing these coefficients as Fourier coefficients of a $2\pi$-periodic 
function $f(\theta)$ according to
\begin{equation}
	b_n = \frac{1}{2\pi} \int_0^{2\pi} \! {\rm d}\theta \, f(\theta)
	{\rm e}^{-{\rm i}(n-r)\theta} \; ,
\end{equation}
this system~(\ref{eq:NNS}) becomes the Mathieu equation
\begin{equation}
	\eta f(\theta) = -\frac{1}{2} E''_r f''(\theta) + 2\hbar\mu_1 
	\langle r | J_z | r - 1 \rangle \cos\theta f(\theta) \; ; 
\label{eq:MEQ}
\end{equation}
substituting $\theta = 2z$ and setting $f(2z) \equiv \chi(z)$ yields
its standard form~\cite{AbramowitzStegun72} 
\begin{equation}
	\left(\frac{\mathrm{d}^2}{\mathrm{d}z^2} + \alpha -2q\cos(2z) \right)
	\chi(z) = 0
\end{equation}
with parameters
\begin{eqnarray}
	\alpha & = & \frac{8\eta}{E''_r} \; , \\ 
	q      & = & \frac{4}{E''_r/(\hbar\omega)} \frac{2\mu_1}{\omega}
	             \langle r |J_z| r - 1 \rangle \; .
\label{eq:MPQ}
\end{eqnarray}

\begin{figure}[t]
\centering\includegraphics[width=1.0\textwidth]{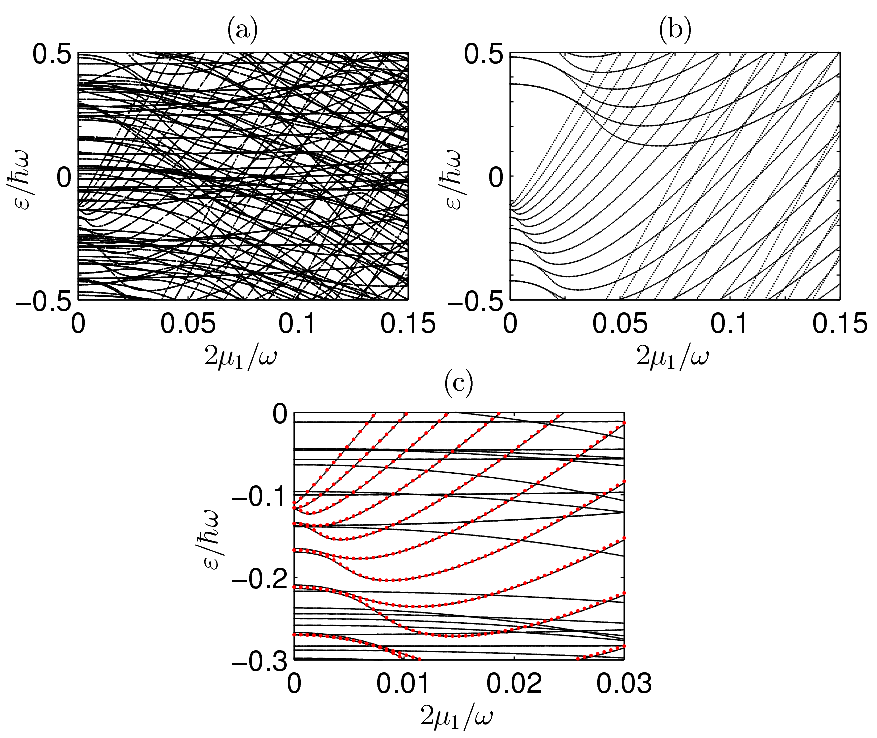}
\caption{Quasienergies for the driven bosonic Josephson junction~(\ref{eq:NPH}) 
	with $N = 100$ particles, scaled interaction strength 
	$N\kappa/\Omega = 1.9$, and scaled driving frequency 
	$\omega/\Omega = 1.6$. 
	(a)~Numerically computed, exact complete spectrum, showing 
	quasienergies of both near-resonant and non-resonant states.	
	(b)~Quasienergies of near-resonant states with $k = 0,\ldots,18$
	according to the approximation~(\ref{eq:MAP}). Here $r = 34$ is the 
	resonant state of the undriven junction; the Mathieu 
	parameter~(\ref{eq:MPQ}) is given by $q\omega/2\mu_1 = 5402$.
	(c)~Exact quasienergies (black lines) compared to the approximate
	ones (red dots) for small scaled driving strengths~$2\mu_1/\omega$.
	Additional lines stem from non-resonant states.}
\label{F_1}	
\end{figure}

The requirement that $\chi(z) = \chi(z+\pi)$ be a $\pi$-periodic Mathieu 
function then restricts~$\alpha$ to one of the discrete characteristic values  
\begin{equation}
	\alpha_k(q) = \begin{cases}
    	a_k(q) 	   & \mathrm{for} \quad k = 0,2,4,\dots \\
    	b_{k+1}(q) & \mathrm{for} \quad k = 1,3,5,\dots \; ,
  	\end{cases}
\label{eq:MCV}
\end{equation}
as tabulated in the mathematical literature~\cite{AbramowitzStegun72}. 
Writing $f_{\ell,k}$ for the $\ell$th Fourier coefficient of the Mathieu
function singled out by $\alpha_k(q)$, one thus has the approximations
\begin{equation}
	| \Psi_k(t) \rangle = 
	\exp\left(-\frac{{\rm i}}{8\hbar} E''_r \alpha_k t \right) 
	\sum_\ell f_{\ell,k} | r + \ell \rangle \exp\left[
	-\frac{{\rm i}}{\hbar}\Big(E_r + \ell\hbar\omega\Big) t \right]
\label{eq:MBF}
\end{equation}
for near-resonant Floquet states; their quasienergies
\begin{equation} 
	\varepsilon_{k} = E_r + \frac{1}{8}E''_r\alpha_k(q) 
	\quad \mathrm{mod}\; \hbar\omega
\label{eq:MAP}
\end{equation}
are determined by the Mathieu characteristic values~(\ref{eq:MCV}). 
In comparison with the exact spectrum depicted in Fig.~\ref{F_1}, computed
numerically by diagonalizing the one-cycle time evolution operator $U(T,0)$, 
this approximation performs quite well for moderate scaled driving strengths 
$2\mu_1/\omega$; the regular fan of quasienergies described by 
Eq.~(\ref{eq:MAP}) is well discernible against a background of further 
quasienergies stemming from non-resonant states. Interestingly, the many-body 
Floquet states~(\ref{eq:MBF}) carry a new quantum number $k = 0,1,2,\ldots\,$, 
but at this point the physical significance of this quantum number does not 
seem to be obvious. In the following section these states~(\ref{eq:MBF}) will 
be interpreted from the mean-field point of view; in particular, it will be 
shown that the state with $k = 0$ has Trojan properties.

\section{$N$-particle -- mean field correspondence}
\label{sec:phase}

In a semiclassical approach, the quantum states of a one-dimensional oscillator
with corresponding classical momentum~$p$ and conjugate coordinate~$x$ are 
characterized by the Bohr-Sommerfeld condition   
\begin{equation} 
	\frac{1}{2\pi}\oint_{\gamma_k} \! p\mathrm{d}x = 
	\hbar\left(k + \frac{1}{2}  \right)
\label{eq:BSC}
\end{equation}
with integer $k = 0,1,2,\ldots\,$, assuming that the classical oscillation 
has two ``soft'' turning points so that the Maslov index of the invariant 
curve $\gamma_k$ in phase space $\{(p,x)\}$ is ${\rm ind} \, \gamma_k = 2$: 
This condition~(\ref{eq:BSC}) singles out those invariant curves from which 
the discrete quantum states are obtained by means of the usual WKB 
construction~\cite{Gutzwiller90}. An appropriately adapted procedure yields 
the Floquet states of a $T$-periodically forced oscillator, provided the 
classical motion is integrable: In this case one has $T$-periodic tubes in the 
extended phase space $\{(p,x,t)\}$ also incorporating the time~$t$ which are 
invariant under the Hamiltonian flow. Then a first quantization condition 
of the form~(\ref{eq:BSC}), with a path $\gamma_k$ winding once around such 
an invariant tube at fixed time~$t$, selects the desired tubes which are 
associated with the Floquet wave functions, whereas a second condition, with 
a $T$-periodic path led along the respective tube, enables one to compute 
their quasienergies~\cite{BreuerHolthaus91}.

\begin{figure}[t]
\centering\includegraphics[width=0.85\textwidth]{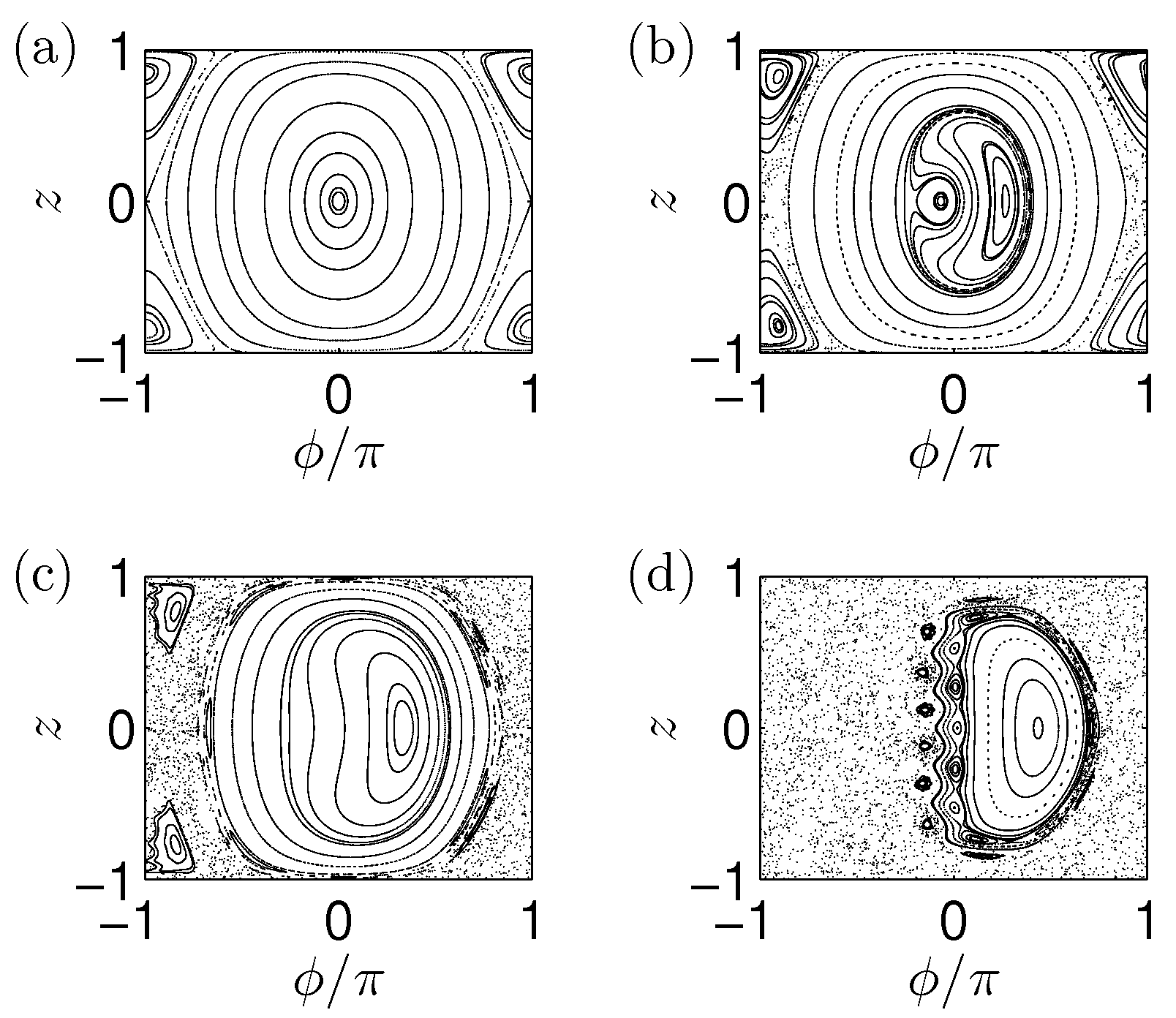}
\caption{Poincar\'{e} surfaces of section for the nonrigid classical
	pendulum~(\ref{eq:NRP}), for $N\kappa/\Omega = 0.95$ and
	$\omega/\Omega = 1.62$. The driving strengh $2\mu_1/\omega$ is 
	given by $0$~(a), $0.02$~(b), $0.1$~(c), and $0.3$~(d). Observe
	the emergence and growth of a resonance zone surrounding a stable
	(elliptic), $2\pi/\omega$-periodic orbit. While the integrability 
	of the unperturbed pendulum is reflected in the regularity of the 
	section~(a), the non-integrability of the driven system manifests
	itself in a chaotic sea emanating from the separatix curve of the 
	undriven motion. In these and all following sections, initial
	conditions are chosen such that all relevant features become visible.}
\label{F_2}	
\end{figure}

One can apply these deliberations to the classical nonlinear driven 
pendulum~(\ref{eq:NRP}), having been obtained as the mean-field approximation
to the $N$-particle system~(\ref{eq:NPH}), and ``re-quantize'' this classical 
pendulum in order to obtain an effective single-particle description. To this 
end, one merely has to observe that the ``momentum'' variable 
$
z = \cos\theta = \cos^2(\theta/2) - \sin^2(\theta/2)
$
directly corresponds, by means of the parametrization~(\ref{eq:SPS}), 
to the operator $2J_z/N$, with $J_z$ as defined by Eq.~(\ref{eq:DJZ}); 
note that $J_z$ actually is the third component of a set of three operators 
$J_x$, $J_y$, $J_z$ satisfying angular-momentum commutation 
relations~\cite{MilburnEtAl97,RaghavanEtAl99}. This implies that the 
standard formula~(\ref{eq:BSC}) here takes the form
\begin{equation} 
	\frac{1}{2\pi}\oint_{\gamma_k} \! z\mathrm{d}\phi = 
	\hbar_{\rm eff}\left(k + \frac{1}{2}  \right)
\label{eq:SQP}	
\end{equation}
with the effective Planck constant
\begin{equation}
	\hbar_{\rm eff} = \frac{2}{N}	
\label{eq:EPC}
\end{equation} 
which equals twice the inverse particle number. However, while the classical
pendulum~(\ref{eq:NRP}) is {\em integrable\/} when $2\mu_1/\omega = 0$, so 
that the semiclassical quantization produces approximations to all its energy 
eigenstates, it becomes non-integrable, and partly {\em chaotic\/}, when the 
driving force is turned on. This is clearly visible in the Poincar\'{e} 
surfaces of section shown in Fig.~\ref{F_2}, obtained as stroboscopic plots 
from exact solutions to the classical equations of 
motion~\cite{ShampineGordon75}. In accordance with the Poincar\'{e}-Birkhoff 
theorem~\cite{Gutzwiller90}, a new stable elliptic fixed point, corresponding 
to a stable $T$-periodic orbit, emerges for weak driving strength when the 
oscillation frequency of the unperturbed pendulum equals the driving frequency;
this fixed point lies in the center of the banana-shaped resonance zone 
appearing in panel~(b). When increasing $2\mu_1/\omega$, this resonance zone 
grows until it becomes an island of mainly regular motion embedded in the 
chaotic sea shown in panel~(d) --- a standard scenario in Hamiltonian systems. 
The invariant curves surrounding the central fixed point of the resonant island
constitute sections of invariant $T$-periodic flow tubes with a plane of 
constant time, and thus provide the quantization paths $\gamma_k$ required by 
Eq.~(\ref{eq:SQP}) for computing semiclassical approximations to the 
$N$-particle Floquet states~\cite{HenkelHolthaus92,BreuerHolthaus91}. Evidently
this procedure now can only yield the ``resonant'' Floquet states carried by 
the island, not covering the states associated with the chaotic sea. However, 
it needs to be stressed that even the apparently regular island actually 
exhibits self-similar chaotic motion on fine scales~\cite{Gutzwiller90}, so 
that here the Bohr-Sommerfeld rule~(\ref{eq:SQP}) applies in a coarse-grained 
sense, glossing over the unresolved details. This will turn out to be important
in Sec.~\ref{sec:floton}.
 
These considerations highlight the particular conceptual value of the 
driven bosonic Josephson junction~(\ref{eq:NPH}): It allows one to invoke 
techniques previously developed in the investigation of the quantum-classical 
correspondence for studying the relation between full quantum $N$-particle 
dynamics and its 
mean-field description~\cite{WeissTeichmann08,GertjerenkenWeiss13}, 
with the large-system limit $N \to \infty$ paralleling, in view of 
Eq.~(\ref{eq:EPC}), the semiclassical limit $\hbar \to 0$.

\begin{figure}[t]
\centering\includegraphics[width=1.0\textwidth]{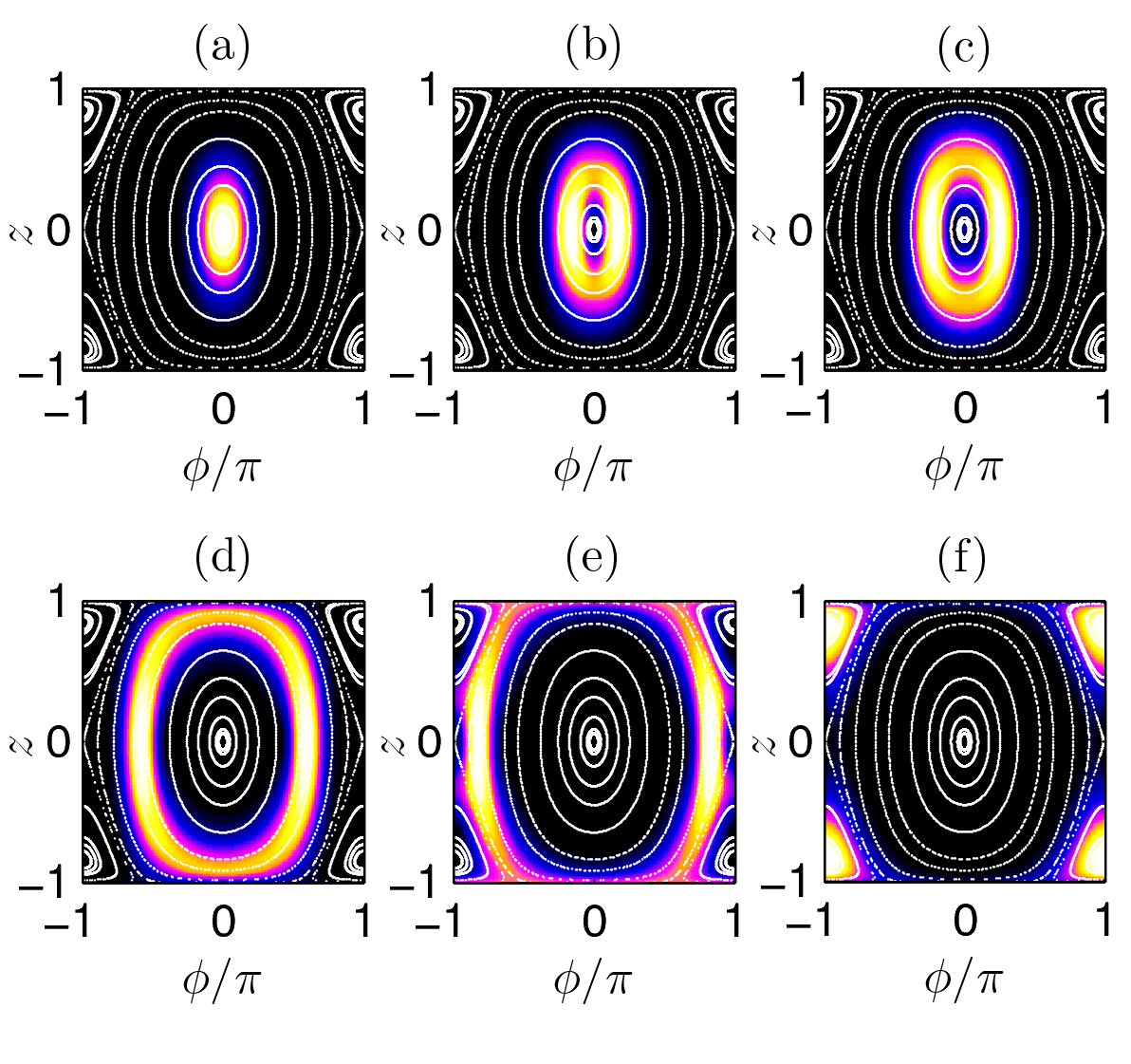}
\caption{Color-coded Husimi distributions~(\ref{eq:HUS}) of exact $k$th 
	$N$-particle energy eigenstates of the undriven bosonic Josephson 
	junction~(\ref{eq:UBJ}) with $N = 20$ and $N\kappa/\Omega = 0.95$,
	superimposed on the corresponding phase-space portrait of 
	the classical nonrigid pendulum~(\ref{eq:NRP}) taken from 
	Fig.~\ref{F_2}(a). Brighter colors amount to larger values of the 
	Husimi distribution. Quantum numbers~$k$  of the states considered 
	are $0,1,2,10,15,20$ [(a)-(f)].}
\label{F_3}
\end{figure}

In the absence of a time-periodic force, that is, for $2\mu_1/\omega = 0$, the 
semiclassical quantization of the nonrigid pendulum~(\ref{eq:NRP}) has been 
explored in significant detail by a large number of authors~\cite{MahmudEtAl05,
GraefeKorsch07,NissenKeeling10,ChuchemEtAl10,SimonStrunz12,GraefeEtAl14}. 
The viability of this approach is underlined by Fig.~\ref{F_3}: Here we take 
$N = 20$ and depict Husimi distributions  
\begin{equation}
	\mathcal{Q}^{(N)}_k(z,\phi) = 
	\left|\langle \theta,\phi | \Psi_k^{(N)} \rangle\right|^2 \; ,
\label{eq:HUS}
\end{equation}
{\em i.e.\/}, squared projections of the exact $k$th $N$-particle energy 
eigenstate $| \Psi_k^{(N)} \rangle$ of the undriven bosonic Josephson 
junction~(\ref{eq:UBJ}) onto the atomic coherent states~(\ref{eq:ACS}), being 
superimposed on the corresponding Poincar\'{e} surface of section of the 
classical pendulum with $z = \cos\theta$. Better than any lengthy verbose
explanation, this figure shows how Eq.~(\ref{eq:SQP}) works: The exact energy 
eigenstates states are semiclassically attached to the invariant curves 
selected by this very condition. As emphasized by Mahmud, Perry, and Reinhard, 
the semiclassical re-quantization gives fairly good quantitative results even 
when the particle number~$N$ is quite small, so that $\hbar_{\rm eff}$ still 
is comparatively large~\cite{MahmudEtAl05}.

\begin{figure}[t]
\centering\includegraphics[width=1.0\textwidth]{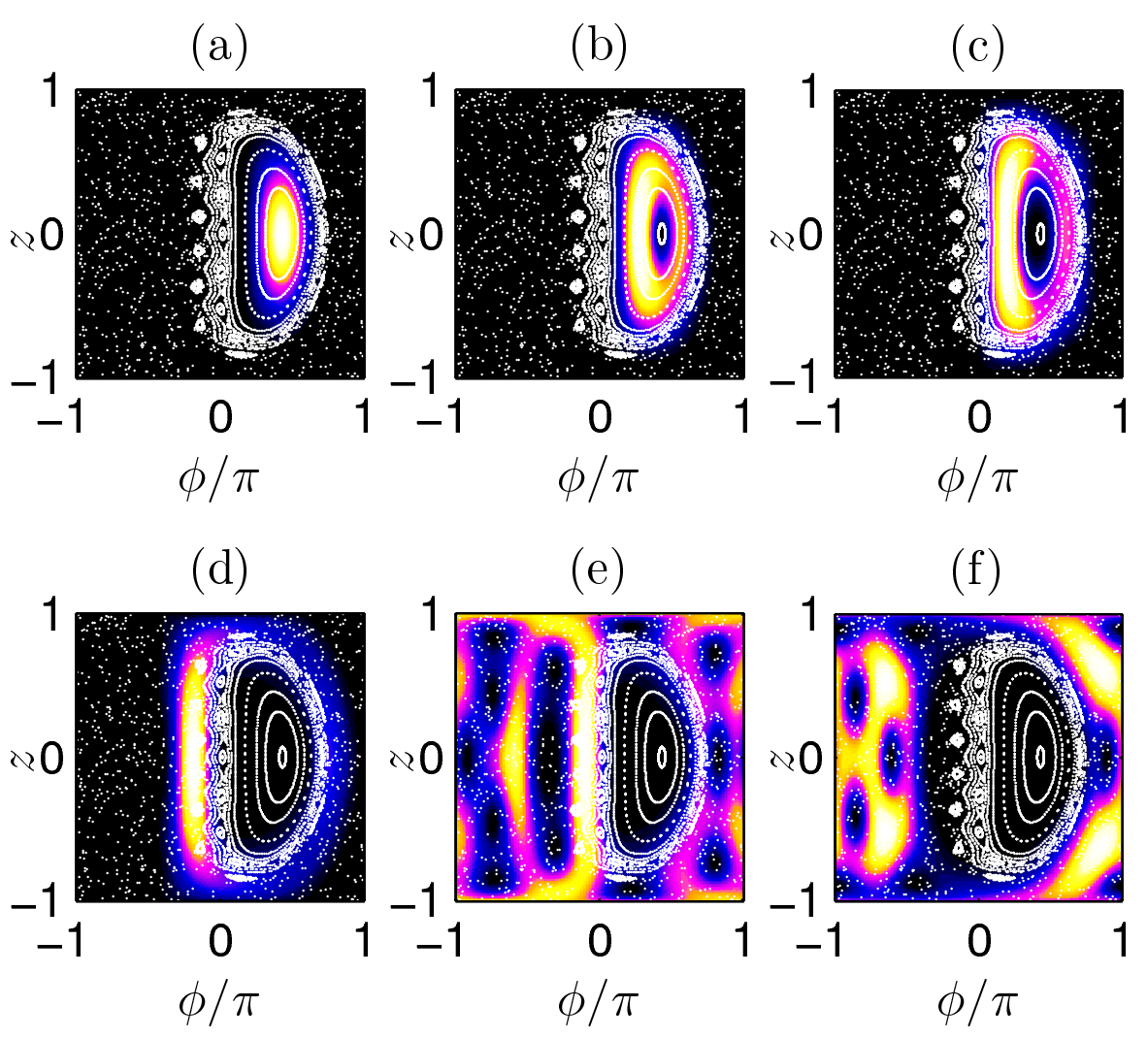}
\caption{Color-coded Husimi distributions~(\ref{eq:HUS}) of exact $k$th
	near-resonant $N$-particle Floquet states of the driven bosonic 
	Josephson junction~(\ref{eq:NPH}) with $N = 20$, 
	$N\kappa/\Omega = 0.95$, $\omega/\Omega = 1.62$, and
	$2\mu_1/\omega = 0.3$, superimposed on the corresponding
	Poincar\'e surface of section taken from Fig.~\ref{F_2}(d).
	Quantum numbers~$k$, determined according to the Mathieu
	approximation~(\ref{eq:MAP}), are $0,1,2,6$ [(a)-(d)]. Panels~(e)
	and~(f) display Husimi distributions of Floquet states associated
	with the stochastic sea.}  	   
\label{F_4}	
\end{figure}

In order to extend this procedure for obtaining a semiclassical understanding
of the Floquet states of the driven bosonic Josephson junction~(\ref{eq:NPH}), 
one only has to compare the Husimi distributions of exact, numerically 
computed $N$-particle Floquet states $| \Psi_k^{(N)}(t_0) \rangle$ with the 
corresponding Poincar\'{e} sections of the driven pendulum~(\ref{eq:NRP}), 
taken at times $t_0 \bmod T$. In Fig.~\ref{F_4} we show such comparisons for 
the conditions of Fig.~\ref{F_2}(d), again taking $N = 20$. Clearly, the 
regular island supports Floquet states which are attached to its invariant 
curves, arising as sections of invariant tubes, exactly as required by the 
quantization condition~(\ref{eq:SQP}). The key point here is that this 
observation leads to a natural explanation of the quantum number~$k$ found in 
Sec.~\ref{sec:resonance} as a result of the Mathieu analysis: The quantum 
numbers~$k$ assigned to the exact near-resonant Floquet states by enumerating
their quasienergies, which are well approximated by the Mathieu
expression~(\ref{eq:MAP}), {\em agree\/} with the quantum numbers~$k$ assigned 
to them on the grounds of the Bohr-Sommerfeld formula~(\ref{eq:SQP}). In 
particular, the ground state-like Floquet state with $k = 0$ is semiclassically
associated with the innermost quantized invariant tube surrounding the stable
$T$-periodic orbit. Precisely the same connection holds in the case of the
nonspreading Rydberg Trojan wave packets~\cite{BialynickiBirulaEtAl94,
KalinskiEberly96}, which again can be semiclassically interpreted as ground
states of a quantized resonance~\cite{HenkelHolthaus92,BuchleitnerEtAl02}.   
Thus, the resonant Floquet state with $k = 0$ of a driven bosonic Josephson
junction constitutes a many-particle counterpart of a Rydberg Trojan state.

\begin{figure}[t]
\centering\includegraphics[width=1.0\textwidth]{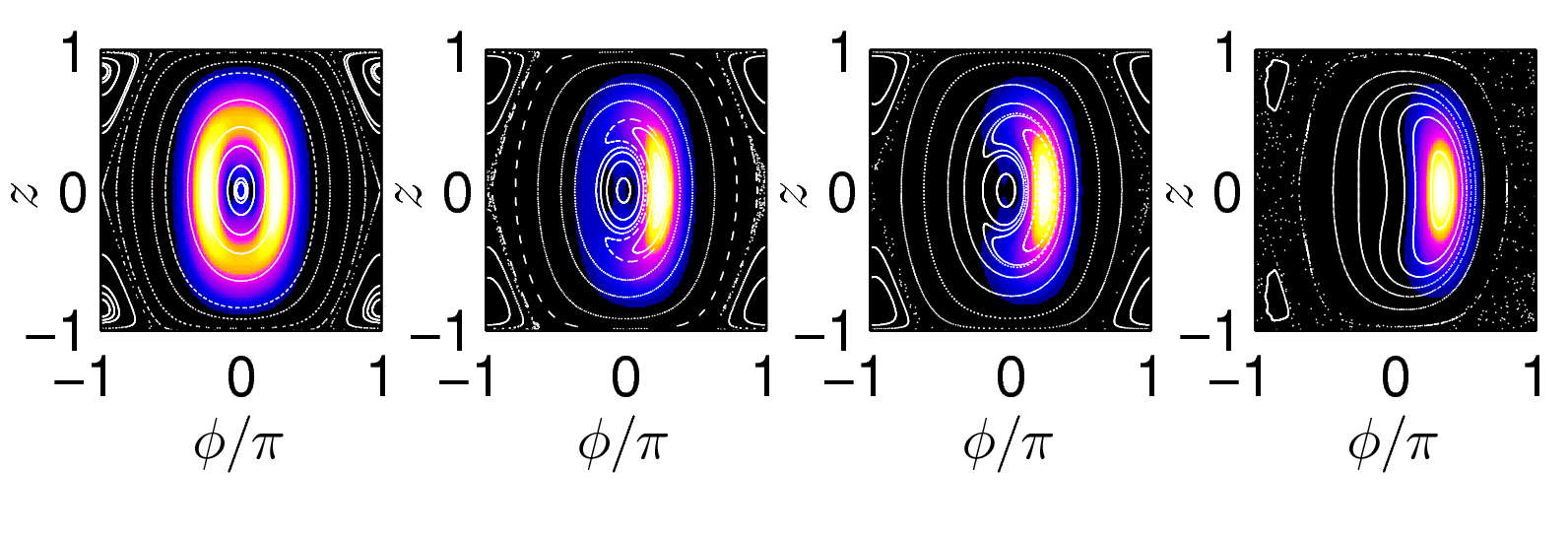}
\caption{Adiabatic transformation of the eigenstate $n = 2$ of the undriven 
	bosonic Josephson junction~(\ref{eq:UBJ}), which is resonant ($2 = r$)
	when $N = 20$, $N\kappa/\Omega = 0.95$, and $\omega/\Omega = 1.62$, 
	into the Trojan Floquet state $k = 0$. From left to right, scaled 
	driving amplitudes $2\mu_1/\omega$ are $0.0$, $0.005$, $0.01$, $0.1$.}
\label{F_5}
\end{figure}

\begin{figure}[htb]
\centering\includegraphics[width=1.0\textwidth]{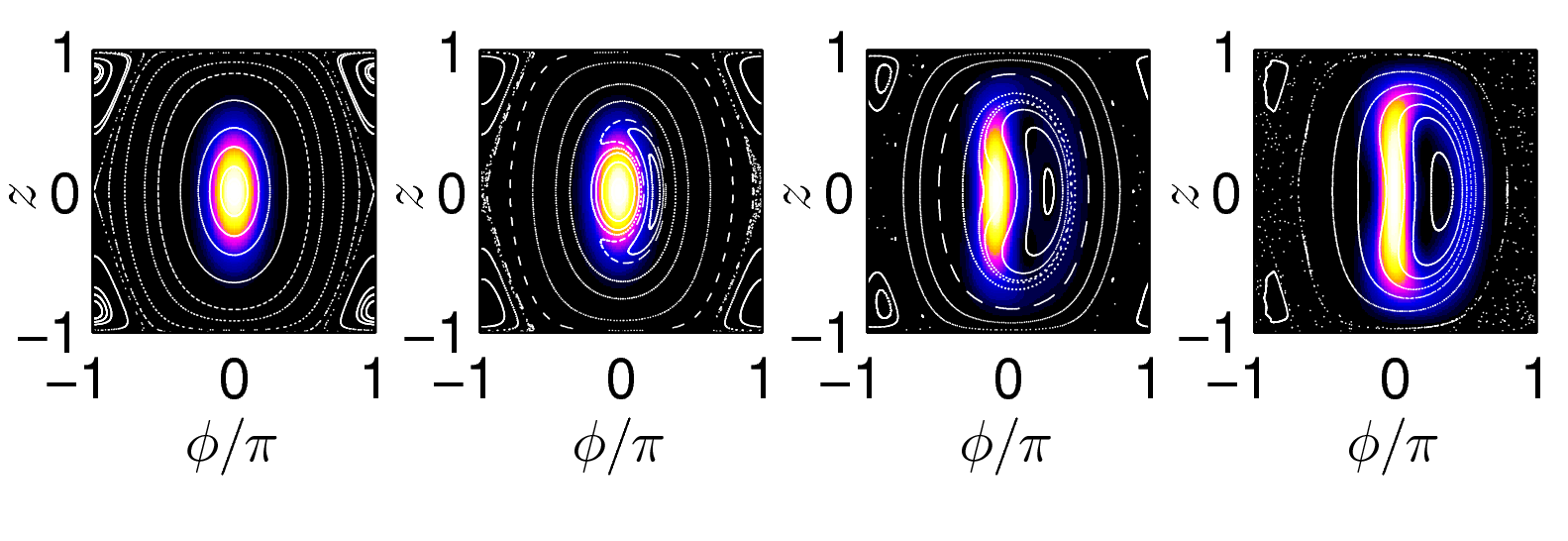}
\caption{Adiabatic transformation of the ground state of the undriven
 	bosonic Josephson junction~(\ref{eq:UBJ}) into a near-resonant
	Floquet state with $k = 2$, for the same parameters as in
	Fig.~\ref{F_5}. From left to right, scaled driving amplitudes 
	$2\mu_1/\omega$ here are $0.0$, $0.005$, $0.04$, $0.1$.}
\label{F_6}	
\end{figure}

These deliberations also imply that the resonant eigenstate $n = r$ of the 
undriven Josephson junction~(\ref{eq:UBJ}) is adiabatically transformed into
the Trojan state $k = 0$ when the external drive is turned on. Fig.~\ref{F_5}
illustrates this metamorphosis for the set of parameters employed so far,
while Fig.~\ref{F_6} shows the reverse transformation of the unperturbed
ground state into an excitation $k > 0$ of the Trojan.

\section{The ``floton'' quasiparticle}
\label{sec:floton}

\begin{figure}[t]
\centering\includegraphics[width=1.0\textwidth]{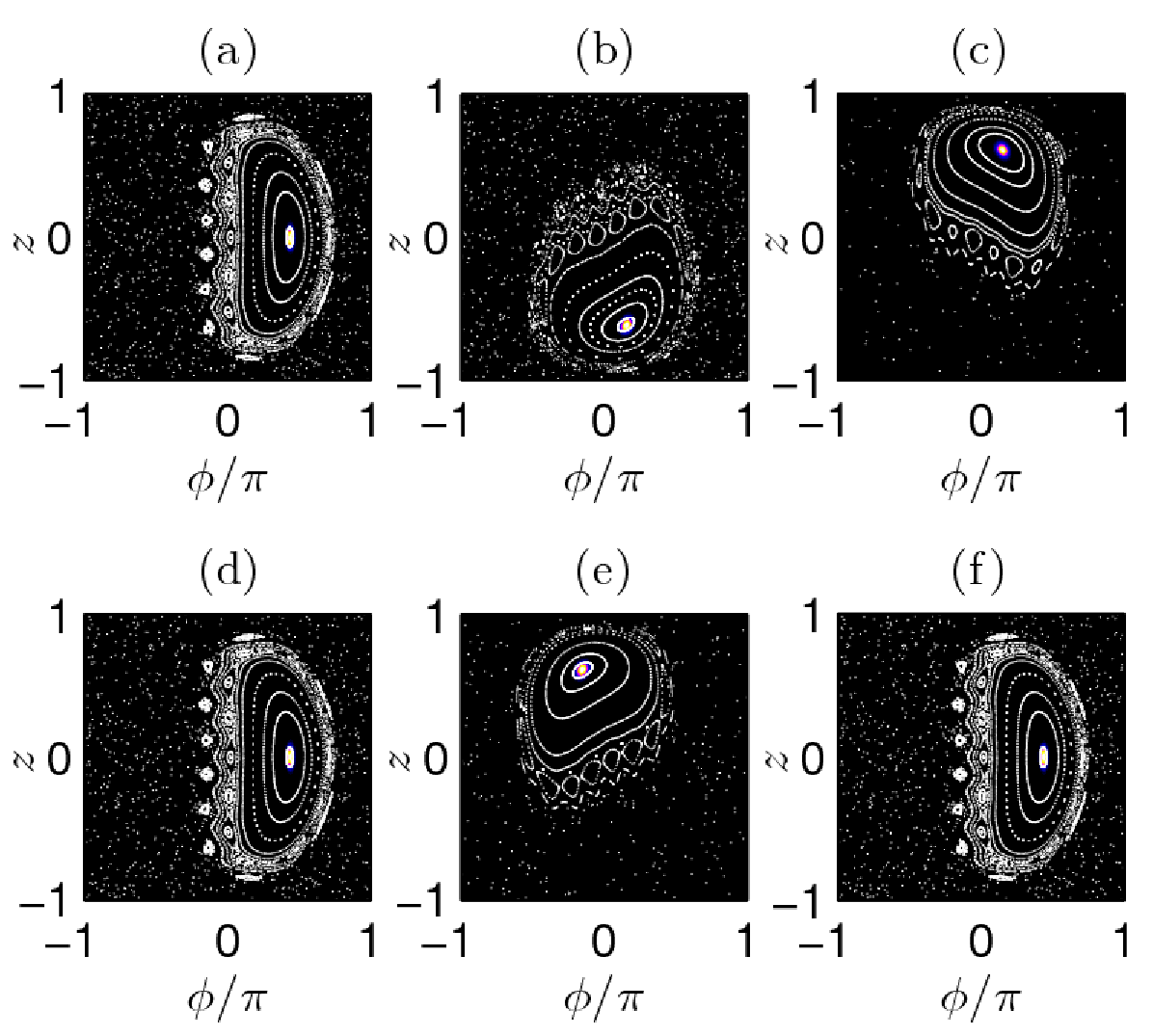}
\caption{Evolution of the Trojan Floquet state~$k = 0$ for $N = 1000$ in 
	time; the other parameters are as in Fig.~\ref{F_4}. Observe that the 
	Husimi distribution is narrowly centered around the periodic orbit, 
	illustrating the scaling implied by the effective Planck constant 
	$\hbar_{\rm eff} = 2/N$. Poincar\'e sections are taken at times~$t/T$ 
	equal to $0.0, 0.2, 0.8, 1.0, 5.7, 10.0$ [(a)-(f)]. This many-body 
	state provides a prototypical example of a floton quasiparticle, 
	being the state with the {\em highest\/} degree of coherence recorded 
	in Fig.~\ref{F_9}. The long-time evolution monitored in panel~(f)
	testifies its periodicity.} 
\label{F_7}
\end{figure}

\begin{figure}[t]
\centering\includegraphics[width=1.0\textwidth]{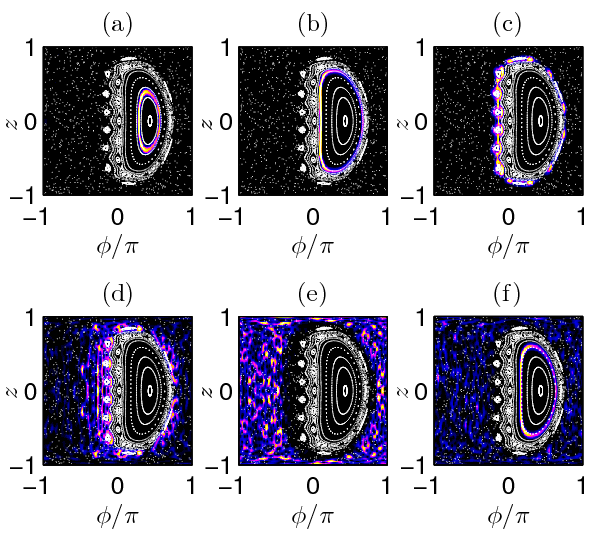}
\caption{Husimi distributions of further exact Floquet states for $N = 1000$;
	the system parameters are the same as in Figs.~\ref{F_4} and~\ref{F_7}.
	Evidently these states are semiclassically associated with the central 
	resonant island~(a,b), with the surrounding chains of higher-order 
	islands~(c,d), or with the chaotic sea~(e). The state depicted in~(f) 
	is a hybrid state, being linked to both an invariant curve within the 
	island and to the chaotic sea; incidentally, this is the state with 
	the {\em lowest\/} degree of coherence recorded in Fig.~\ref{F_9}.}
\label{F_8}
\end{figure}

The small particle number $N = 20$ had been chosen in the previous section 
mainly for illustrative purposes, but with respect to topical laboratory 
experiments~\cite{GatiOberthaler07} particle numbers on the order of $1000$ 
are more to the point. Therefore, Fig.~\ref{F_7} shows the evolution of a 
Trojan with $N = 1000$ in time, again under the conditions of the Poincar\'{e} 
section displayed in Fig.~\ref{F_2}(d). Here the scaling dictated by the
effective Planck constant~(\ref{eq:EPC}) manifests itself in a striking
manner: In comparison with Fig.~\ref{F_4}, where $\hbar_{\rm eff} = 0.1$, one
now has $\hbar_{\rm eff} = 0.002$, implying that the Trojan Floquet state 
$k = 0$ here is associated with a flow tube much closer to the elliptic
periodic orbit than it was in Fig.~\ref{F_4}, so that its Husimi distribution 
appears to be sharply centered around this orbit. By the same token, the 
resonant island now can carry more near-resonant Floquet states with $k > 0$, 
and the Floquet states can resolve much finer details of phase space; this 
is illustrated by the Husimi distributions of further representative Floquet 
states, both near-resonant and non-resonant ones, collected in Fig.~\ref{F_8}.

The Trojan state displayed in Fig.~\ref{F_7} represents an exact collective 
response mode of Bose particles in the driven double-well potential which is 
quite particular in several respects. Firstly, it is ``nonspreading in phase 
space'', being semiclassically attached to a $T$-periodic tube remaining 
invariant under the Hamiltonian flow, and therefore it does not heat up in the 
course of time despite the action of the time-periodic drive. Secondly, it 
behaves like a single particle --- namely, the fictitious particle trapped in 
the ground state $k = 0$ described by the Mathieu equation~(\ref{eq:MEQ}), 
which is an effective Schr\"odinger equation for a (quasi-)particle in a cosine
well with periodic boundary conditions, having led to the Floquet-Mathieu
approximation~(\ref{eq:MBF}). Such quasiparticles have been named ``flotons'' 
in Ref.~\cite{Holthaus95}. In this sense, the many-particle analog of a 
nonspreading Rydberg Trojan wave packet is a nonheating floton quasiparticle; 
this quasiparticle possesses a well-defined discrete set of excited states 
with $k > 0$.  

While most of the arguments employed in the previous section are nothing 
but immediate adaptions of tools routinely used in the discussion of the 
correspondence between quantum and classical systems~\cite{Gutzwiller90}, 
there is one feature which does not occur in that well-established field, but 
deserves particular attention. In our case the classical system~(\ref{eq:NRP}) 
is not given ``as such'', but emerges only as a mean-field description of a 
quantum many-particle system. The reduction of the latter to the mean-field 
level does not constitute a systematic approximation, but involves the 
uncontrolled factorization of expectation values of products of operators into 
products of expectation values of individual operators~\cite{Leggett01}. Thus, 
it cannot be taken for granted that the solution of the mean-field equations 
of motion actually ``corresponds'' to the dynamics of the $N$-particle system. 
Indeed, as discussed in technical detail by Castin and Dum, one expects 
depletion of the condensate when the solution to the mean-field equations of 
motion tends to become chaotic~\cite{CastinDum97}. Essentially, the reduction
of the full $N$-particle dynamics, here being given in terms of vectors with
$N+1$ complex components, to the mean-field level with its two-component
order parameter is viable only if the $N$-particle original is sufficiently
simple, or ordered, in the sense that it has the form~(\ref{eq:ACS}) of an 
$N$-fold occupied single-particle state at least to a good aproximation.
Otherwise the descent to the mean-field level is thwarted by a drastic
loss of information --- if there is no order, there is no order parameter.
Under conditions such that the $N$-particle quantum system does not admit 
the introduction of an order parameter, one can of course still solve the
mean-field equations of motion, but the solutions acquire a meaning which is
different from those cases where an order parameter actually does exist.

Therefore, it is necessary to quantify the ``simplicity'' of an $N$-particle
quantum system, amounting to its ``degree of mean-field approximability'',
or coherence. Such information is contained in the one-particle reduced density
matrix~$\varrho$, which, in the particular case of our two-mode model, adopts 
the form 
\begin{equation}
	\varrho = \left( \begin{array}{cc}
	\langle a_1^{\dagger} a_1^{\phantom{\dagger}} \rangle &
	\langle a_1^{\dagger} a_2^{\phantom{\dagger}} \rangle \\
	\langle a_2^{\dagger} a_1^{\phantom{\dagger}} \rangle &
	\langle a_2^{\dagger} a_2^{\phantom{\dagger}} \rangle 
		\end{array} \right) \; ,
\end{equation}		
the expectation values being taken with respect to the quantum state under
investigation. Obviously the invariant trace of $\varrho$, which is the sum 
of its two eigenvalues, equals the total particle number~$N$. If the larger 
eigenvalue alone already is close to~$N$, then $\varrho$ approaches a 
projection operator, times $N$, onto the associated eigenvector. This indicates
that there exists a simple condensate, {\em i.e.\/}, an almost $N$-fold 
occupied single-particle state constituting the order parameter. In the other 
extreme where both eigenvalues are close to $N/2$ the condensate is fragmented. 
Hence, Leggett has introduced the ``degree of simplicity'' (coherence)   
\begin{equation}
	\eta  =2 N^{-2} \, {\rm tr} \, \varrho^2 - 1
\label{eq:ETA}
\end{equation}	
which is computed from the trace of the squared single-particle density 
matrix~\cite{Leggett01}: Its maximum value $\eta = 1$ indicates a perfect
condensate guaranteeing optimal mean-field approximability, whereas the 
minimum value $\eta = 0$ signals maximal fragmentation.

\begin{figure}[t]
\centering\includegraphics[width=1.0\textwidth]{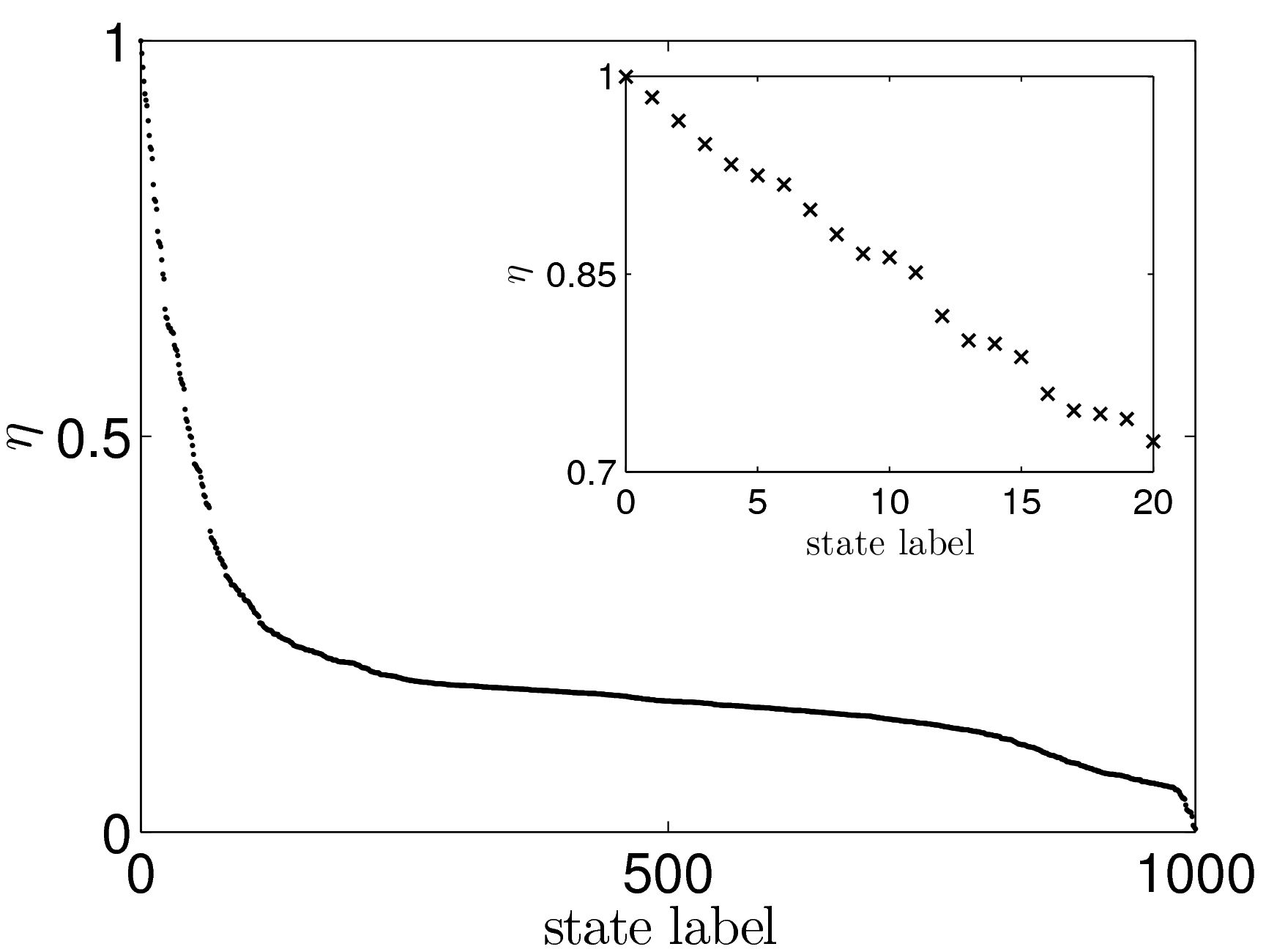}
\caption{Degree of coherence~$\eta$, as defined by Eq.~(\ref{eq:ETA}), 
	for all 1001 Floquet states furnished by the driven Josephson 
	junction~(\ref{eq:NPH}) with parameters as in Figs.~\ref{F_7} 
	and~\ref{F_8}. States are ordered with respect to decreasing
	magnitude of $\eta$. The highest value $\eta_{\rm max} = 0.9997$, 
	signaling close-to-perfect mean-field approximability, is attained 
	by the floton inspected in Fig.~\ref{F_7}; the lowest value 
	$\eta_{\rm min} = 0.0038$ is provided by the hybrid state shown 
	in Fig.~\ref{F_8}(f).} 
\label{F_9}
\end{figure}

It is, therefore, of interest to inspect the simplicity~$\eta$ of the Floquet
states provided by the driven Josephson junction~(\ref{eq:NPH}), in order 
to assess the relevance of the chaotic classical mean-field solutions. In 
Fig.~\ref{F_9} we plot $\eta$ for all~$1001$ Floquet states which arise for 
the parameters of Figs.~\ref{F_7} and \ref{F_8}. This calculation reveals the 
third characteristic feature of the floton: Under ideal conditions, as  met 
here, it gives rise to a value of $\eta$ which is quite close to one, and thus 
possesses close-to-perfect mean-field approximability. The excited states 
with $k > 0$ carried by the resonant island are responsible for the rapidly 
decreasing leftmost part of the plot, while the states mainly associated with 
the chaotic sea lead to the extended plateau; some hybrid states, such as the 
one exemplified by Fig.~\ref{F_8}(f), carry a particularly low value of $\eta$.
   
It is with regard to this pertinent problem of mean-field approximability 
that the further investigation of many-body Trojan states, {\em i.e.\/}, of 
flotons, may bear some significance. Since such states can be generated by 
adiabatic following, as witnessed by Fig.~\ref{F_5}, they should be 
experimentally accessible. One key question then is to what extent the features
extracted from our model system do survive in experimentally realistic set-ups,
which in general will be way too sophisticated to admit full $N$-particle 
modeling, but for which mean-field calculations may still be feasible. Here 
the knowledge that floton-like mean-field solutions do indeed have a faithful 
$N$-particle counterpart would be quite valuable. However, there is still 
another catch. Namely, with increasing particle number~$N$ and correspondingly 
decreasing effective Planck constant $\hbar_{\rm eff} = 2/N$ the system is 
able to explore its phase space on ever finer scales, and ultimately ``feels'' 
that the resonant island does not represent perfectly integrable dynamics, 
but rather is subject to the Kolmogorov-Arnold-Moser (KAM) scenario: The 
invariant flow tubes with not sufficiently irrational winding numbers are 
destroyed~\cite{Gutzwiller90}. But this means that, in a strict mathematical 
sense, the question arises whether or not the quantization curves $\gamma_k$
required by the naive Bohr-Sommerfeld condition~(\ref{eq:SQP}) actually are 
available --- if not, the order parameter should be degraded. Since the answer 
to this question depends sensitively on the precise value of $\hbar_{\rm eff}$,
and therefore on the precise particle number~$N$, one has to expect strong 
fluctuations of the system's coherence under small variations of~$N$, once 
$N$ becomes sufficiently large.

\begin{figure}[t]
\centering\includegraphics[width=1.0\textwidth]{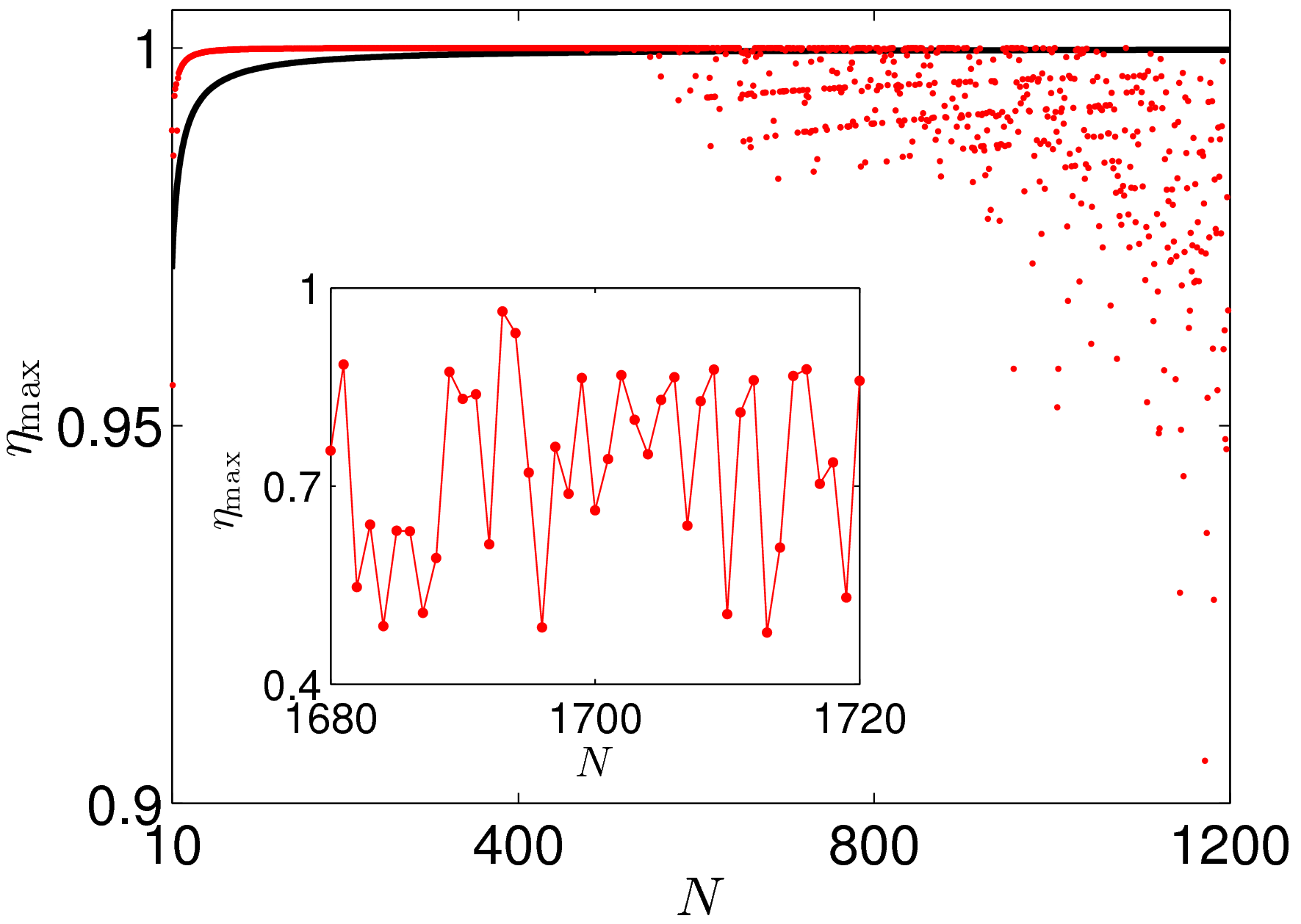}
\caption{Red dots: Maximum degree of coherence $\eta_{\rm max}$, taken over 
	all respective Floquet states, as function of the particle number~$N$. 
	Once again, the system parameters are $N\kappa/\Omega = 0.95$, 
	$\omega/\Omega = 1.62$, and $2\mu_1/\omega = 0.3$. For comparison,
	the black line shows $\eta_{\max}$ for the nondriven junction.
	Observe the scale of the inset!}
\label{F_10}
\end{figure}

This expectation is confirmed in an impressive manner by Fig.~\ref{F_10}, 
which shows the respective maximum $\eta_{\rm max}$ of the $\eta$-values of all 
Floquet states provided by the driven junction~(\ref{eq:NPH}) for a prescribed 
value of~$N$, again for the same set of system parameters as employed before. 
Indeed, at about $N \approx 500$ fluctuations of $\eta_{\rm max}$ start to make
themselves felt, and become more pronounced with increasing $N$; the previous 
observation that $\eta_{\rm max}$ is close to unity for $N = 1000$ now appears 
as a fortuitous coincidence. In contrast, such fluctuations do not occur for 
the undriven junction~(\ref{eq:UBJ}), which merely gives rise to exactly 
integrable mean-field dynamics.  

Such fluctuations of the order parameter constitute the fourth, and probably
most important property of the floton, and lead to a hard, testable prediction:
In a set of experiments with resonantly driven Bose-Einstein condensates, 
prepared under ostensibly identical conditions but still admitting a small
uncertainty of the large particle number, there should be large shot-to-shot
fluctuations of the condensate fraction.

\section{Experimental implications}
\label{sec:final}

Quantum resonances, as sketched in Sec.~\ref{sec:resonance}, constitute a 
generic feature of driven nonlinear systems. Therefore, the appearance of 
Trojan quasiparticles which correspond, to good approximation, to an $N$-fold
occupied time-periodic single-particle state is not restricted to the somewhat 
idealized model system~(\ref{eq:NPH}). In particular, one may deliberately 
``engineer'' such a resonance in an anharmonic single-well trapping potential 
which is combined with additional time-periodic forcing. A central question 
then is how such time-dependent condensates could be prepared. Here a further 
insight becomes important: As emphasized by Leggett~\cite{Leggett01}, the 
tendency to undergo Bose-Einstein condensation into a ``macroscopically'' 
occupied single-particle state is not restricted to states of thermal 
equilibrium. Rather, it can be understood as a consequence of two mutually 
reinforcing effects: On the one hand, bosonic configurations in which many 
particles occupy the same single-particle orbital have a higher statistical 
weight than in the classical case; on the other, the Hartree-Fock energy of 
two identical spinless bosons in different single-particle orbitals is greater 
than that of two such bosons in the same orbital~\cite{Leggett01}. Hence, one 
may reasonably assume that a resonantly driven Bose gas condenses ``by itself''
into the floton ground state $k = 0$ when the parameters are chosen such
that this Floquet state becomes the one with the lowest mean energy. If so, 
the floton would manifest itself, {\em e.g.\/}, as a clearly discernible 
condensate peak in a time-of-flight recording, at least as long as the particle
number is not too large. Because of the nonheating property, this signature 
should persist even after comparatively long driving times. But the true 
``smoking gun'' ultimately betraying the floton might consist in fluctuations 
of the kind depicted in Fig.~\ref{F_10}: If the average particle number 
exceeds a certain value depending on the specific set-up, the height of 
the condensate peak observed in a series of time-of-flight experiments would 
fluctuate in a seemingly erratic manner from measurement to measurement.

\begin{acknowledgments}
We acknowledge discussions with C.~Weiss.
This work was supported by the Deutsche Forschungsgemeinschaft (DFG) under 
grant no.\ HO~1771/6-2. 
The computations were performed on the HPC cluster HERO, located at the 
University of Oldenburg and funded by the DFG through its Major Research 
Instrumentation Programme (INST 184/108-1 FUGG), and by the Ministry of 
Science and Culture (MWK) of the Lower Saxony State.
\end{acknowledgments}

\end{document}